\PassOptionsToPackage{numbers,sort&compress}{natbib}  
\documentclass[preprint,12pt]{elsarticle}
\usepackage[margin=2.5cm]{geometry}
\usepackage{enumitem}
\newlist{inlinelist}{enumerate*}{1}
\setlist*[inlinelist,1]{%
  label=(\roman*),
}
\usepackage{dirtytalk}
\usepackage{soul}
\usepackage{hhline}
\usepackage{subfig}
\usepackage{multirow}
\usepackage{xcolor}
\usepackage{graphicx}
\usepackage{upgreek}
\usepackage{amssymb}
\usepackage{amsfonts,amsthm,bm,amsmath} 
\usepackage{appendix}
\usepackage{times}
\usepackage{caption}
\usepackage{subfig}
\usepackage{tablefootnote}
\newcommand{\psubref}[1]{\protect\subref{#1}}
\usepackage{multirow}
\usepackage[noabbrev]{cleveref}

\newcommand{\fref}[1]{Fig.~\ref{#1}}

\newcommand{\eref}[1]{Eq.~(\ref{#1})}

\newcommand{\sref}[1]{Section~\ref{#1}}
\newcommand{\tref}[1]{Table~\ref{#1}}
\setcitestyle{square}

\begin{document}

\begin{frontmatter}

\title{Designing impact-resistant bio-inspired low-porosity structures using neural networks}
\author[]{Shashank Kushwaha$^1$}
\author[]{Junyan He$^1$}
\author[]{Diab Abueidda$^2$}
\author[]{Iwona Jasiuk$^1$\corref{mycorrespondingauthor}}
\address{$^1$ Department of Mechanical Science and Engineering, University of Illinois at Urbana-Champaign, Urbana, IL, USA \\
$^2$ National Center for Supercomputing Applications, University of Illinois at Urbana-Champaign, Urbana, IL, USA}
\cortext[mycorrespondingauthor]{Corresponding author}
\ead{ijasiuk@illinois.edu}
\begin{abstract}
Biological structural designs in nature, like hoof walls, horns, and antlers, can be used as inspiration for generating structures with excellent mechanical properties. A common theme in these designs is the small percent porosity in the structure ranging from 1 - 5\%. In this work, the sheep horn was used as an inspiration due to its higher toughness when loaded in the radial direction compared to the longitudinal direction. Under dynamic transverse compression, we investigated the structure-property relations in low porosity structures characterized by their two-dimensional (2D) cross-sections. A diverse design space was created by combining polygonal tubules with different numbers of sides placed on a grid with varying numbers of rows and columns. The volume fraction and the orientation angle of the tubules were also varied. The finite element (FE) method was used with a rate-dependent elastoplastic material model to generate the stress-strain curves under plane strain conditions. A gated recurrent unit (GRU) model was trained to predict the structures' stress-strain response and energy absorption under different strain rates and applied strains. The parameter-based model uses eight discrete parameters to characterize the design space and as inputs to the model. The trained GRU model can efficiently predict the response of a new design in as little as 0.16 ms and allows rapid performance evaluation of 128000 designs in the design space. The GRU predictions identified high-performance structures, and four design trends that affect the specific energy absorption were extracted and discussed.

\section*{\bf{Graphical abstract}}
{\centering
\includegraphics[width=0.95\textwidth]{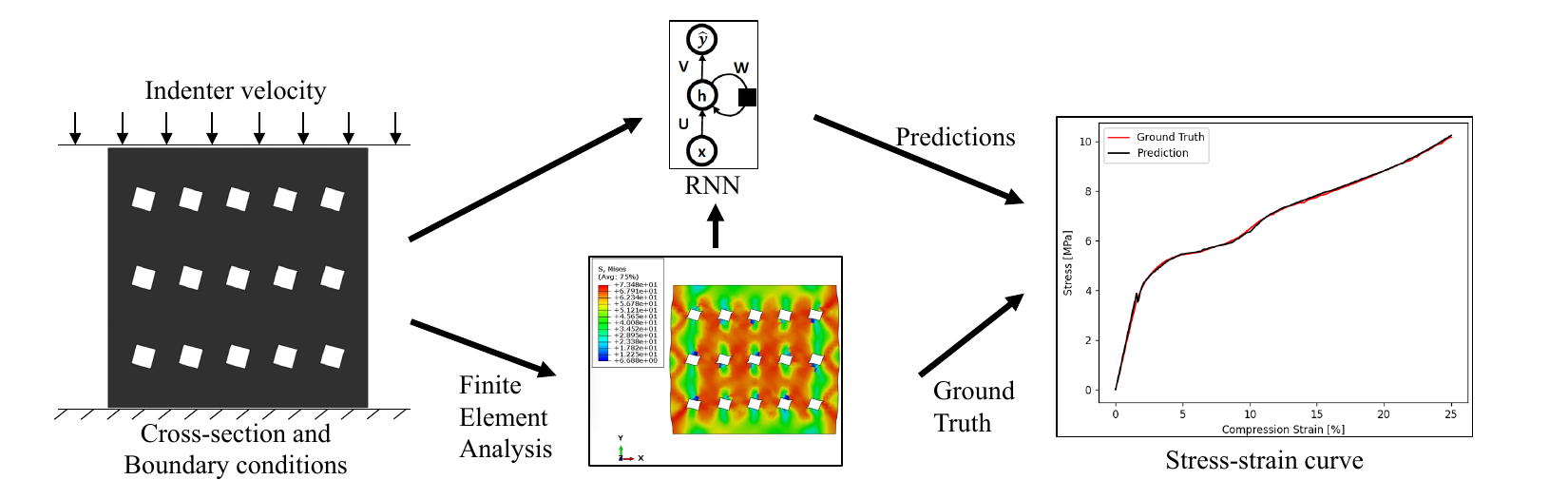}
\par
}
\end{abstract}

\begin{keyword}
Bio-Inspired \sep Structure-property relations \sep Neural networks \sep Specific energy absorption 
\end{keyword}

\end{frontmatter}

\section{Introduction}
\label{sec:intro}
Lightweight structures with high energy absorption capacity are of high interest for multiple engineering applications. Various structural elements found in animals and plants could be used as inspiration to design novel structures that can sustain impacts generated during collision \citep{HA2020107496, LAZARUS202015705, MEYERS20081}. The process of evolution has created complex architectures in nature capable of handling low-to-medium velocity impacts (up to 50 m/s). An example is the trabecular-honeycomb biomimetic structure inspired by beetle elytra \citep{yu_pan_chen_zhang_wei_2018}. Rams see impact velocities of around 5.5 m/s when fighting. Also, during collisions, sheep horns can withstand a maximum impact force of 3400N \citep{kitchener_1988}. The sheep microstructure has evolved to sustain large dynamic forces without catastrophic failure \citep{drake2016horn}. Similarly, the hoof sustains high impact loading forces close to 9000 N while galloping \citep{bertram1986fracture}. The tubular structure is a common feature in equine hoofs and horns \citep{kasapi1997design, tombolato2010microstructure}. Such structure contains arrays of long aligned tubules within the bulk material, promoting energy absorption.

Biological materials and structures often exhibit excellent energy absorption capabilities and inspire the design of new energy absorbers. Bio-inspired structures have been used in countless applications, including automobiles \citep{GONG2020107081}, protective armors \citep{bioinspired_armor}, and wings of aircraft \citep{MISHRA2015519}. Further, a variety of materials have been used to manufacture bio-inspired structures, including polymers \citep{Zhang2017}, aluminum alloy \citep{GONG2020107081}, fiber-reinforced composites \citep{AMORIM2021109336}, and concrete \citep{TOADER2017369}. Hence, studying the structure-property relations of bio-inspired designs is of great research and industrial interest. The exploration of structure-property relations involves surveying many different structural features at a given loading condition. Various studies utilized optimization-based methods to generate new designs for energy absorption and study the structure-property relations \citep{duddeck2016topology,he2022latticeopt, zeng2017improved, sharafi2014shape, he2023size}. However, a systematic compilation of bio-inspired designs’ mechanical response and energy absorption characteristics is lacking. In previous studies, the response of the hoof and horn-inspired structures was studied at quasi-static loading \citep{islam2022quasi}. Various types of designs, including but not limited to composite laminates \citep{islam2022quasi, rice2019horse} and tubular honeycomb structures \citep{ma2022crushing}, have been tested using experiments and finite element analyses (FEA) \citep{ingrole2021bioinspired}. The primary objective of these evaluations was to obtain greater energy absorption or damage tolerance through crack deflection. Further, it was shown by Sabet et al. \citep{sabet2018mechanical} that the geometrical arrangement of stiff and soft phases can significantly influence the overall properties of the composite structure.

Within the solid mechanics domain, neural network (NN) models have been extensively used to predict stress-strain response of composites \citep{yang2020prediction, chen2019machine, ABUEIDDA2019111264}, metals \citep{gangi2014artificial,HE2023116277,he2023deep}, and lattices \citep{laban2020experimental, 10.1115/1.4045040, hassanin2020controlling}. However, the use of NN models for studying bio-inspired structures remains scarce. Existing studies have utilized GANs to design porous structures using X-ray microtomography images as input \citep{siegkas2022generating}. Apart from GANs, bio-inspired structures have been designed using a conditional variational autoencoder \citep{chiu2023designing}. In most cases, either a specific property \citep{ding2023accelerating} is predicted, or in an unsupervised deep learning method, images or parameters of the structure are predicted \citep{shen2022nature}. Previous works did not focus on predicting the full-field temporal distribution of the stress field during the impact.

Thus, the prediction of stress fields as a function of time is the first objective of this study. Further, this paper aims to develop a systematic framework to generate structures that combine different design elements found in low-porosity structures in nature, i.e., the study of the structures with aligned tubules whose porosity is in the range of 1\% - 5\% under transverse dynamic compression. The framework generates low-porosity structures with constant cross-sections along the thickness direction by randomly combining various design features such as tubule shape, orientation, and in-plane arrangement. Once trained, the NN can efficiently predict the mechanical performance of new designs at a rate much faster than classical numerical simulations, thus allowing rapid preliminary design selection and trend identification. Therefore, the second objective of this work is to develop a neural network (NN) model to approximate the structure-property relations, linking the input design parameters with loading conditions and the mechanical performance of the structure. Structure-property maps of the design space at different loading rates are identified, and design trends are discussed.

This paper is organized as follows. \sref{sec:method} presents an overview of the numerical simulations, the input data preprocessing, and the NN model's architecture. \sref{sec:results} includes the results obtained from the study and explores the quality of NN predictions and the validity of the results. \sref{sec:conc} summarizes the outcomes and lists some possible future directions for the bio-inspired structures.

\section{Methods}
\label{sec:method}
\subsection{Geometry generation and Finite element analysis}
\label{sec:fe_sim}
\begin{figure}[h!]
\begin{center}
    \includegraphics[scale=0.5]{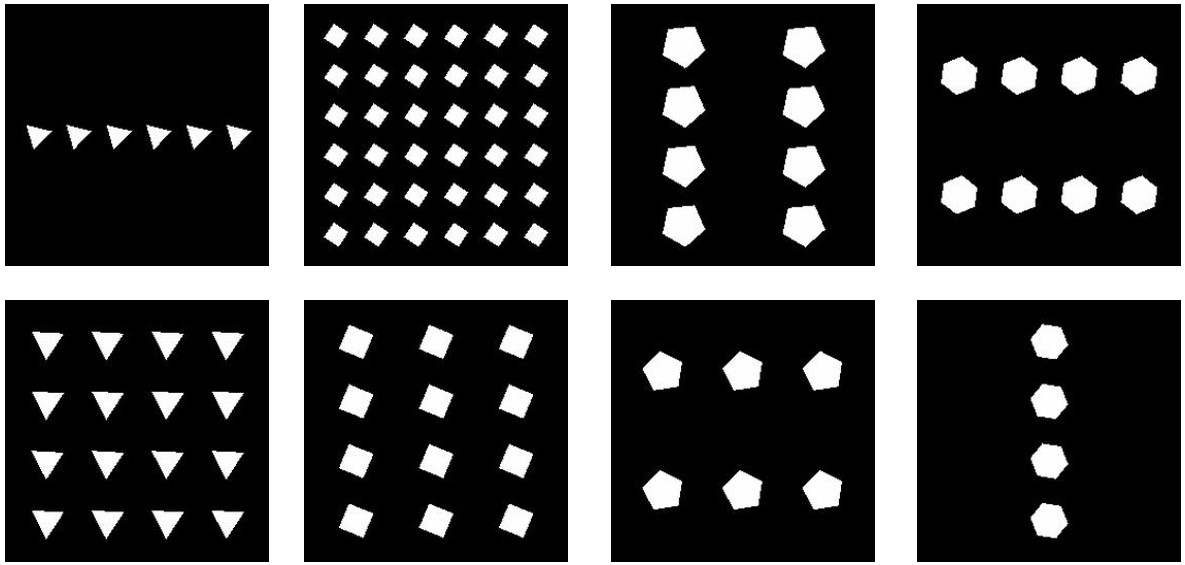} 
    \caption{Sample structures in the design space}
    \label{sample_designs}
\end{center}
\end{figure}

The designs considered in this work are 3D structures containing tubules with a constant cross-section. Hence, the designs can be uniquely characterized by their 2D, in-plane cross-sections, assuming the plane strain condition. A Python script was developed to generate cross-sectional sketches in the finite element (FE) analysis package Abaqus \citep{Abaqus2021} for a given volume fraction, tubule shape, tubule orientation, and the arrangement of the tubules within the structure. The cross-section of the bio-inspired structures studied in this work is an 11-by-11 mm$^{2}$ square, whereas all the tubules are confined within a concentric square area of 10-by-10 mm$^{2}$. The tubule volume fraction was uniformly sampled from the range [1\%, 10\%]. In this work, we approximated the tubule cross-sections by polygons of a different number of sides that were uniformly sampled from the range [3, 6] i.e., included triangles, squares, pentagons, and hexagons. Additionally, rotation was applied to the cross-sections, and the rotation angle was uniformly sampled from the range [0, 360] degrees. Multiple tubules can be present in the structure, and we placed them on a $n_y \times n_x$ grid, where $n_y$ and $n_x$ denote the number of rows and columns, respectively. However, all the tubules in a given configuration have the same shape, and the designs with non-intersecting tubules were considered valid. Other designs were excluded from the analysis.The $n_y$ and $n_x$ were sampled in the range [1, 8]. Some selected structures in the design space are shown in \fref{sample_designs}. All the structures were discretized with 4-node bilinear plane strain quadrilateral elements with reduced integration. A nominal element edge length of 0.24 mm was chosen for meshing. 

The relationship between different structural designs and energy absorption mechanisms seen in bones, teeth, and horns is discussed by McKittrick et al.\citep{MCKITTRICK2010331}. Further, they discuss that when rams butt heads, the horns are loaded in the transverse direction, which provides more energy absorption than in the longitudinal direction. 

\begin{figure}[h!]
\begin{center}
    \includegraphics[scale=0.7]{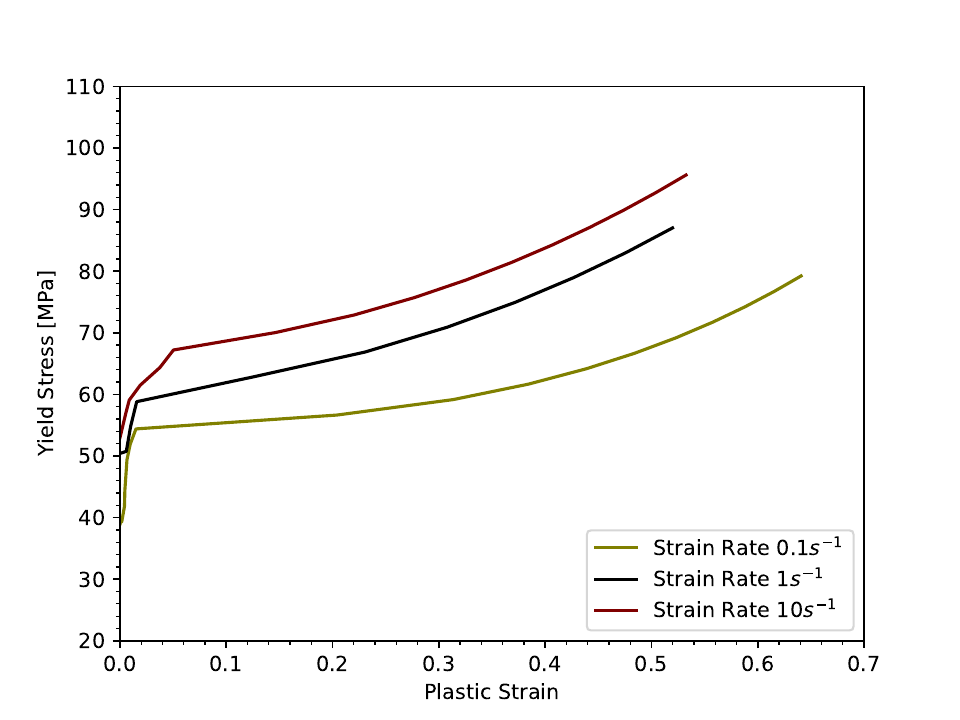} 
    \caption{Yield Stress versus Plastic Strain at different strain rates}
    \label{material_model}
\end{center}
\end{figure}

The Abaqus/Explicit dynamic simulation defined a rate-dependent elastic-plastic material model to capture the structures' response at varying strain rates. The strain rate decomposition is given by \citep{Abaqus2021}: 

\begin{equation}
\label{1}
d\boldsymbol{\epsilon} = d\boldsymbol{\epsilon}^{el} + d\boldsymbol{\epsilon}^{pl}
\end{equation}
Using the definition of corotational measures, the integrated form is given by \citep{Abaqus2021}: 

\begin{equation}
\label{2}
\boldsymbol{\epsilon} = \boldsymbol{\epsilon}^{el} + \boldsymbol{\epsilon}^{pl}
\end{equation}
The elasticity is linear and isotropic defined using Young's modulus, E, and Poisson's ratio, $\nu$. The flow rule is \citep{Abaqus2021}:

\begin{equation}
\label{3}
d\boldsymbol{e} = d\bar{e}^{\boldsymbol{pl}} \boldsymbol{n}
\end{equation}
where
\begin{equation}
\label{4}
\boldsymbol{n} = \frac{3}{2} \frac{\boldsymbol{S}}{q} 
\end{equation}
\begin{equation}
\label{5}
q = \sqrt{\frac{3}{2} \boldsymbol{S}:\boldsymbol{S}} 
\end{equation}
and $d\bar{e}^{\boldsymbol{pl}}$ is the (scalar) equivalent plastic strain rate. The plasticity required that the material satisfy a uniaxial-stress plastic-strain-rate relationship. In case of rate dependence, the uniaxial flow rate is defined as follows \citep{Abaqus2021}: 
\begin{equation}
\label{6}
\boldsymbol{\dot{\bar{e}}^{pl}} = \boldsymbol{h}(q, \bar{e}^{\boldsymbol{pl}}, \theta)
\end{equation}
where $\bar{e}^{\boldsymbol{pl}}$ is the equivalent plastic strain, $\theta$ is the temperature, and $\boldsymbol{h}$ is a known function. The overstress power law model in the rate-dependent material model is defined as follows \citep{Abaqus2021}: 
\begin{equation}
\label{7}
\boldsymbol{\dot{\bar{e}}^{pl}} = D \left( \frac{q}{\sigma_{0}} - 1  \right)^{n} 
\end{equation}
where $D(\theta)$ and $n(\theta)$ are user defined temperature-dependent material parameters and $\sigma_{0}(\bar{e}^{pl}, \theta)$ is the static yield stress. Integrating \eref{7} by the backward Euler method gives:
\begin{equation}
\label{8}
\Delta\boldsymbol{\bar{e}}^{pl} = \Delta t \boldsymbol{h}(q, \bar{e}^{\boldsymbol{pl}}, \theta) 
\end{equation}
\eref{8} can be inverted to obtain $q$ as a function of $\bar{e}^{pl}$ at the end of the increment. Hence, the uniaxial form is given by \citep{Abaqus2021}: 
\begin{equation}
\label{9}
q = \bar{\sigma}(\bar{e}^{pl}) 
\end{equation}
where $\bar{\sigma}$ is obtained by inverting \eref{8}. \Cref{1,2,3,4,5,6,7,8,9} are used to define material behavior. At every increment when the plastic flow is occurring, these equations are integrated and solved for the state at the end of the increment. 
The material properties of the base material chosen for the study are similar to polycarbonate-acrylonitrile butadiene styrene (PC-ABS). The Young's modulus and Poisson's ratio are 2.5 GPa and 0.35, respectively. The strain rate dependent yield stress versus plastic strain curves used to define the plastic region are included in \fref{material_model}. However, the strains to failure are tremendous in horns, as much as 80\% \citep{MCKITTRICK2010331, zhang2018microstructure}. The structures considered in this study have low porosity. At large nominal strains, most of the porosities would already be compressed. Consequently, the stress response primarily arises from the material's densification. This perspective is further reinforced by the absence of damage modeling in our study. Additionally, conducting FE simulations up to high nominal strain would demand considerably more time for input data generation for the neural network. Hence, the maximum nominal strain considered is 25\%.

\begin{figure}[h!] 
    \centering
     \subfloat[]{
         \includegraphics[clip,width=0.45\textwidth]{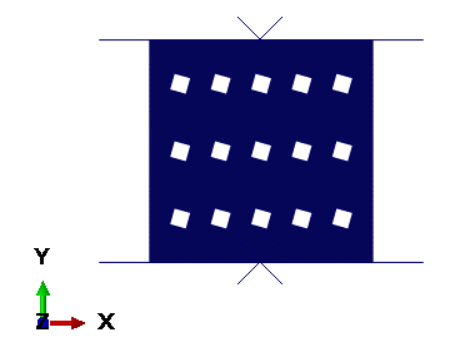}
         \label{fig:cae}
     }
     \subfloat[]{
         \includegraphics[clip,width=0.55\textwidth]{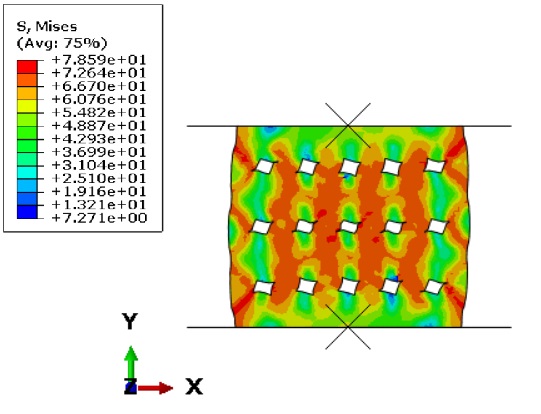}
         \label{fig:mises}
     }
    \caption{FE model setup and results: \psubref{fig:cae} Typical structure with two rigid plates for dynamic transverse compression. 
    \psubref{fig:mises} A typical deformed structure showing von Mises stress.}
    \label{cae}
\end{figure}

In this study, the boundary conditions for impact loading were approximated by sandwiching the structure between two rigid plates, and the structures were subjected to dynamic transverse compression. The bottom plate was held fixed, and the top plate traveled downward with a constant velocity determined by the user-defined strain rate. The nominal strain rate was uniformly sampled from the range [0.45, 90.9] s$^{-1}$ corresponding to indenter velocity from the range [5, 1000] m/s.  The reaction force and displacement were measured at the top rigid plate. All sidewalls were traction-free and were free to deform. All simulations had a constant final displacement of 2.25 mm, corresponding to 25\% nominal compressive strain along the y-axis. The reaction force and displacement at the top plate, plastic dissipation, and elastic strain energy of the porous structures were outputs of the FE simulations. \fref{cae} depicts the FE model assembly and a typical deformed structure at the end of dynamic compression.
A total of 7196 simulations were conducted on an AMD Ryzen 7 5800H processor with 8 cores. Depending on the applied impact velocity, each simulation took about 5-30 minutes to complete. 

\subsection{Neural network for sequence prediction}
\label{sec:NN}

\subsubsection{Input data, data augmentation, and loss function}
\label{sec:data_loss}
The input parameter range is described in \sref{sec:fe_sim}. The corresponding output arrays were obtained from the impact simulations conducted in Abaqus/Explicit. The output arrays were down-sampled to 50-time steps for the efficiency of neural network training. The inputs used in the model consist of eight temporal information arrays. The first five arrays are constant in time and correspond to the parameters used to define the structure's geometry. The parameters include ${n}$: the topology of the tubule (i.e., number of sides in a polygon), ${n}_{x}$: number of tubules evenly distributed in the x-direction, ${n}_{ y}$: number of tubules evenly distributed in the y-direction, ${a}_{ o}$: rotation angle for all the tubules in the structure, and ${v}_{ f}$: volume fraction of the individual tubule in each element created by ${n}_{ x}$ times ${n}_{ y}$ elements in a 10-by-10 mm$^{2}$ grid. The remaining three inputs are physics-informed temporal arrays described as follows: 
\begin{enumerate}
    \item Current time value at each output time point.
    \item Nominal compression strain at each output time point.
    \item Nominal compression strain rate.
\end{enumerate}

A standard scaler in Scikit-Learn normalized all the inputs \citep{scikit-learn} before training. The scaler was fitted only to the training data points to avoid information leakage \citep{yang2020prediction}. The available training data was increased using data augmentation. Corresponding to each simulation conducted in Abaqus with 25\% final nominal strain, twenty final nominal strains in the range [10\%,25\%] were randomly sampled, and all inputs and outputs were linearly interpolated to the selected final strain level. This method generated training data points at the same strain rate but different final nominal strain, and increased the total number of input data points from 7196 to 719600. These data points were divided into training (65\%), validation (15\%), and testing datasets (20\%). 

The mean absolute error (MAE) has been employed as the loss function in this study \citep{willmott2005advantages}. The loss function is defined as:
\begin{equation}
    {\rm{MAE}} = \frac{ \sum^N_{i=1} |\bf{Y}_i - \hat{\bf{Y}}_i| }{ N },
\end{equation}
where $N,\bf{Y}_i,\hat{\bf{Y}}_i$ denote the number of training data points, ground-truth outputs, and the NN predictions, respectively. The mean squared error (MSE) is  chosen as a metric, which is defined as:
\begin{equation}
    {\rm{MSE}} = \frac{ \sum^N_{i=1} (\bf{Y}_i - \hat{\bf{Y}}_i)^2 }{ N }.
\end{equation}

\subsubsection{Neural network model}
\label{sec:nn_model}
This study uses a recurrent neural network (RNN) model to train the forward model for output prediction. Specifically, the gated recurrent unit (GRU) model is used. This model has been widely used to predict sequences \citep{he2023exploring, abueidda2021deep, frankel2019predicting,he2023sequential}. Further, Abueidda et al. \citep{abueidda2021deep} compared the performance of different RNN models to predict the response of elastoplastic material undergoing deformation under variable strain rates. Although the GRU model is more computationally expensive than the long short-term memory (LSTM) model and the temporal convolutional network (TCN) model, it predicts the output with lower error. Based on the GRU model's demonstrated capabilities to predict the structures' response under complex deformation histories, this study used the model to predict stress-strain curves for the structures under dynamic transverse compression. 

\begin{figure}[h!]
\begin{center}
    \includegraphics[trim={0 1.5cm 6cm 0cm},width=\textwidth]{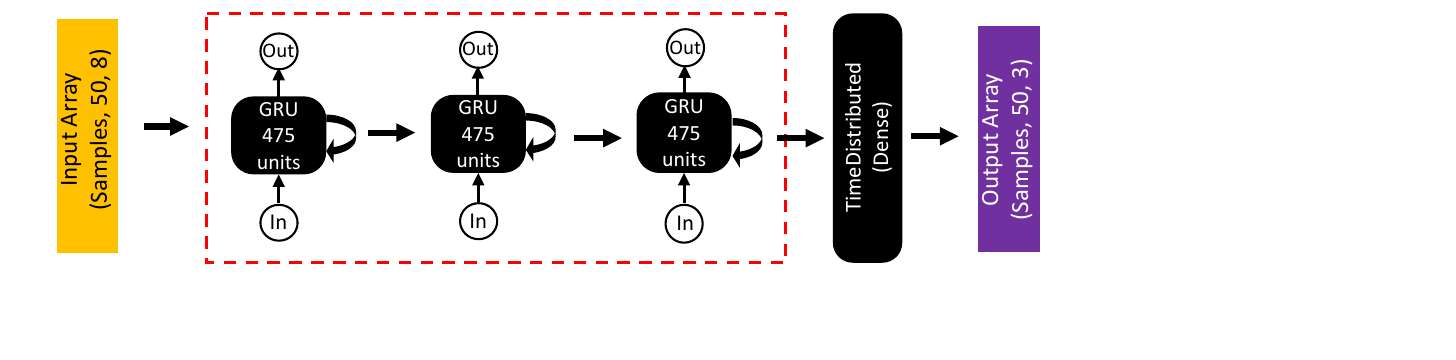} 
    \caption{Neural network architecture}
    \label{NN_architecture}
\end{center}
\end{figure}

The GRU-based model was implemented and tested in Keras \citep{chollet2015keras} with a TensorFlow \citep{tensorflow2015-whitepaper} backend. The GRU model comprises three stacked layers of 475 GRU units, each with hyperbolic tangent (tanh) activation leading to a model with 3.77 million trainable parameters. The NN architecture is presented in \fref{NN_architecture}. The loss function was minimized using an Adam optimizer \citep{kingma2014adam} with an initial learning rate of 1×10$^{-3}$. The model was trained for 150 epochs with a batch size of 600, and training was repeated 10 times to obtain average training time and model accuracy. The data set was shuffled and partitioned in each training repetition, as described in \sref{sec:data_loss}. All training was conducted on Google Colab Pro+ using GPU acceleration on Tesla V100 GPU.

\subsection{Global optimization}
\label{sec:opt}
Using the trained neural network, a Python script was developed to traverse the input design space and evaluate the energy absorption performance. The input design space was divided into grid points based on the first five input parameters described in \sref{sec:data_loss}. Each grid point represents a unique structure within the input design space based on five input parameters. The specific energy absorption (SEA) was computed for each grid point by calculating the area under the load-displacement curve (calculated from the GRU model predictions). Three design parameters: number of sides of the polygon, ${n}_{x}$ and ${n}_{y}$ could take discrete integer values within their respective input range, whereas volume fraction and angle offset were divided into 40 and 20 equally spaced intervals, respectively. Hence, this method was used to analyze the SEA for 128000 structures within the input design space. This process was repeated for five different values of the indenter velocity (${v}_{y}$) within the range described in \sref{sec:method}. A similar process can be repeated at different equally spaced intervals to obtain the performance of all the structures in the input design space for a given final strain and the indenter velocity.

\section{Results and discussion}
\label{sec:results}

\subsection{Validation of the neural network predictions}
\label{sec:SEA_validate}
\begin{figure}[h!] 
    \centering
     \subfloat[]{
         \includegraphics[trim={8cm 4cm 8cm 4cm},clip,width=0.23\textwidth]{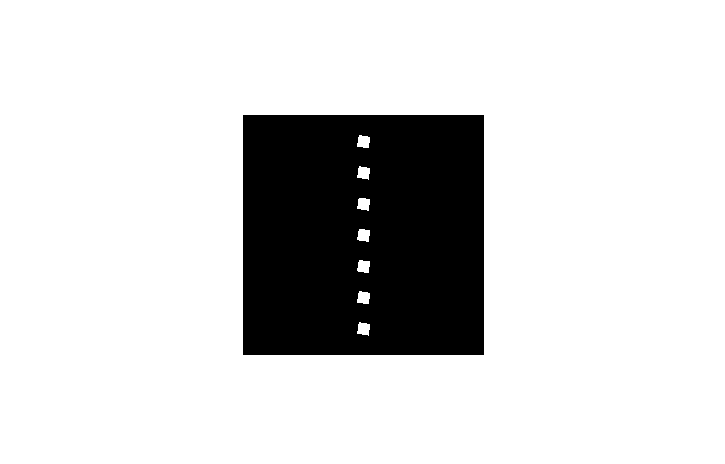}
         \label{i2}
     }
     \subfloat[]{
         \includegraphics[trim={8cm 4cm 8cm 4cm},clip,width=0.23\textwidth]{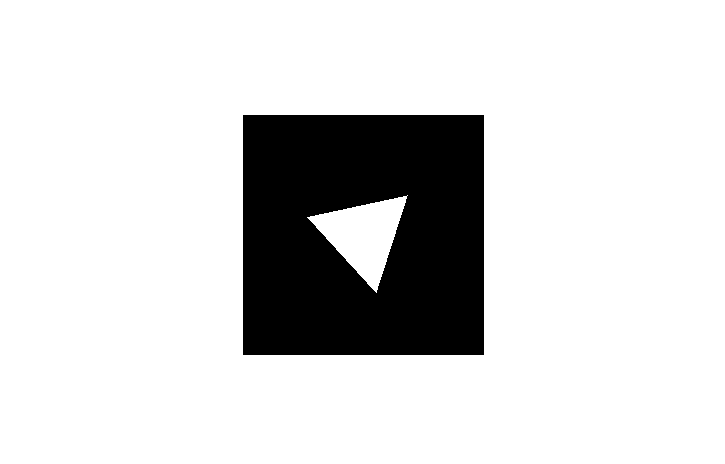}
         \label{i3}
     }
     \subfloat[]{
         \includegraphics[trim={8cm 4cm 8cm 4cm},clip,width=0.23\textwidth]{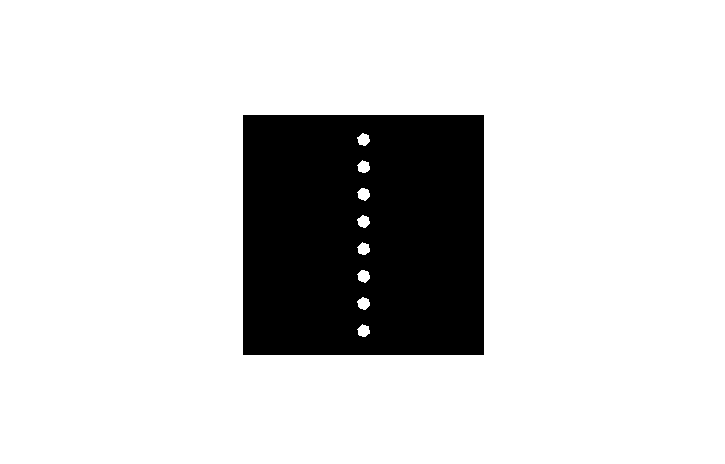}
         \label{i0}
     }
     \subfloat[]{
         \includegraphics[trim={8cm 4cm 8cm 4cm},clip,width=0.23\textwidth]{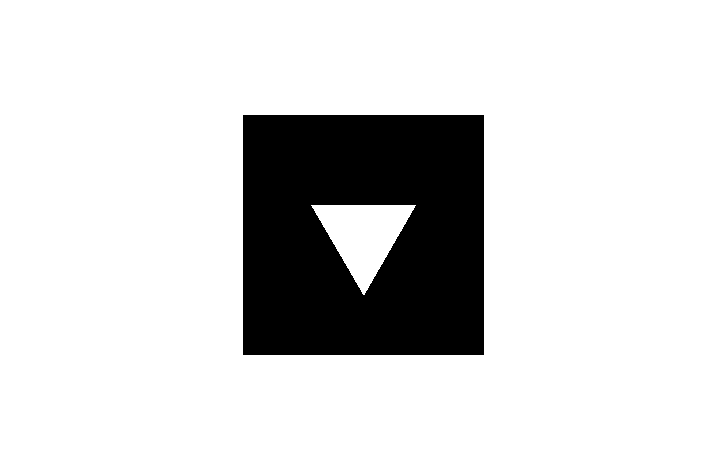}
         \label{i1}
     }
    \caption{Highest and lowest SEA designs as predicted by the trained GRU model: \psubref{i2} highest SEA, 10 m/s, \psubref{i3} lowest SEA, 10 m/s, \psubref{i0} highest SEA, 100 m/s, \psubref{i1} lowest SEA, 100 m/s.}
    \label{validatation_designs}
    \centering
         \includegraphics[width=0.4\textwidth]{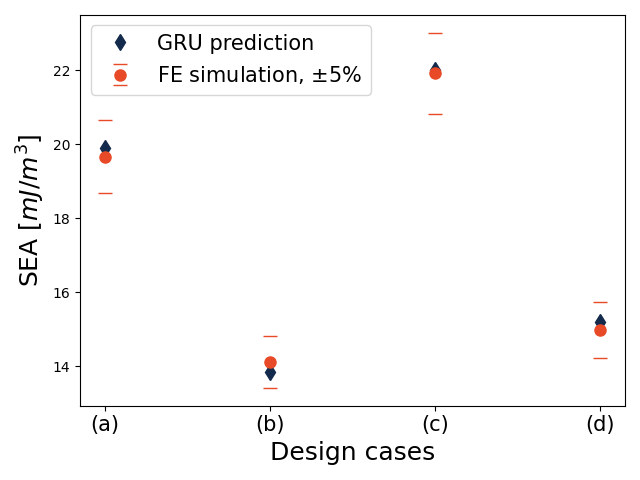}
    \caption{Comparison of the FE-simulated and GRU-predicted SEA values for the four validation design cases.}
    \label{validation_sea}
\end{figure}

The best and the worst designs (as predicted by the trained GRU model) at two different impact velocities (10 and 100 $m/s$) were validated by FE simulations to check the accuracy of the GRU model predictions. \fref{validatation_designs} shows the best and the worst designs at two different indenter velocities, specifically 10m/s and 100m/s. 
FE simulations were conducted to obtain the ground-truth values of SEA under an applied plate velocity of 10 $m/s$ (cases (a) and (b)) and 100 $m/s$ (cases (c) and (d)) and a final axial strain of 0.25. The comparison of the FE-simulated and GRU-predicted SEA values are shown in \fref{validation_sea}. As can be seen from the results, the trained GRU is highly accurate for the two impact velocities tested, and the predicted SEA values fall within 5\% of their respective ground truth values. This result provides confidence in applying the trained model for further inference tasks.

\subsection{Predicting stress-strain curves and energy outputs}
\label{sec:energy}
\begin{figure}[h!] 
    \newcommand\x{0.5}
    \newcommand\y{5cm}
    \centering
     \subfloat[]{
         \includegraphics[trim={0 0 0cm 0cm},clip,width=\x\textwidth, height=\y]{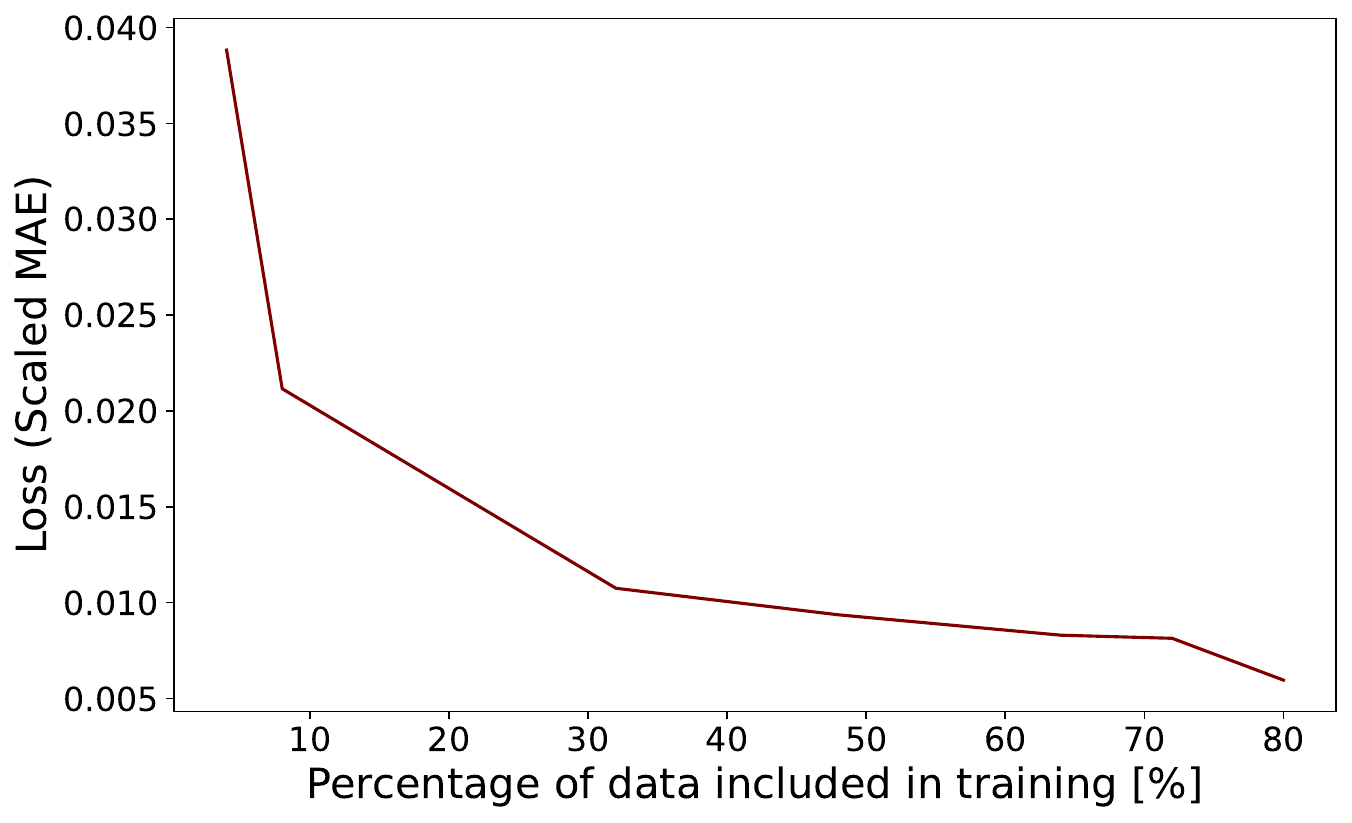}
         \label{fig:pct_plot}
     }
     \subfloat[]{
         \includegraphics[trim={0 0 0cm -0.1cm},clip,width=\x\textwidth, height=5.05cm]{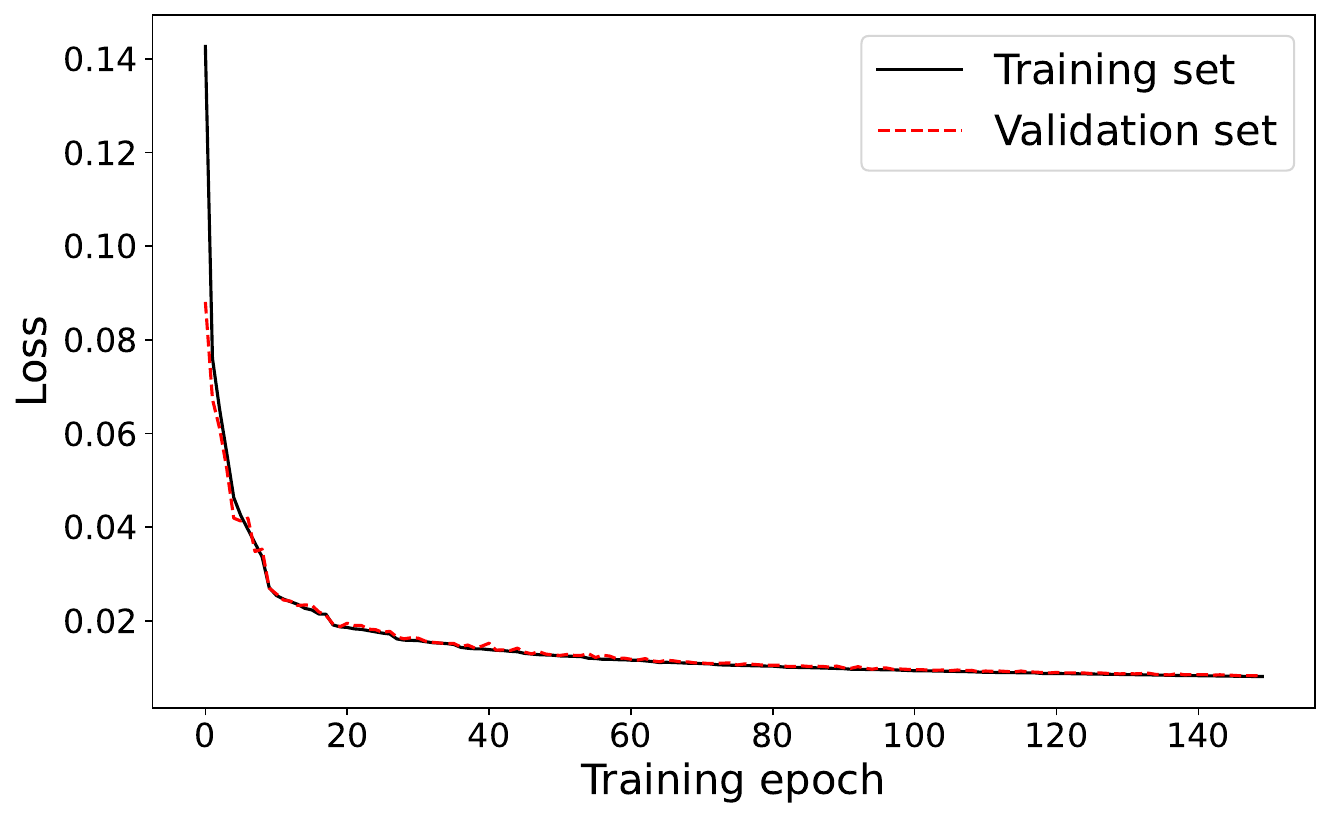}
         \label{fig:loss}
     }
    \caption{Convergence plot for GRU model training process: \psubref{fig:pct_plot} Scaled mean squared error when a different percentage of the total data is used in training. \psubref{fig:loss} Scaled mean absolute error evolution during training. Note that the MAE shown here is the MAE computed on the variables scaled by the standard scaler.}
    \label{conv}
\end{figure}

The number of input data points used in training was decided based on the prediction accuracy measured using the value of the loss function. In this study, the percentage of total input data was incremented to train the neural network model until similar prediction accuracy was observed. Further, the average response of the GRU model was measured by training the model 10 times after shuffling the data before each training iteration. The loss function value corresponding to the increasing amount of training data is shown in \fref{fig:pct_plot}. Further, a typical training history is also presented in \fref{fig:loss}. The average training and inference times for the GRU model and the average FE simulation time are reported in \tref{time_comp}.
\begin{table}[h!]
    \caption{Computational cost for GRU training, inference, and FE simulations}
    \small
    \centering
    \begin{tabular}{ccccccc}
      & \vline & GRU training & GRU inference & FE simulation \\
    \hline
    Time & \vline  & 5192.9s & 1.63$\times 10^{-4}$s  & 5-30 mins \tablefootnote{Depends on the impact velocity of the rigid plate. A lower impact speed leads to a longer solution time due to the small time step size used in the explicit analysis.} \\
    \end{tabular}
    \label{time_comp}
\end{table}

\begin{figure}[h!]
\begin{center}
    \includegraphics[trim={0cm 0cm 0cm 0cm},clip,width=\textwidth]{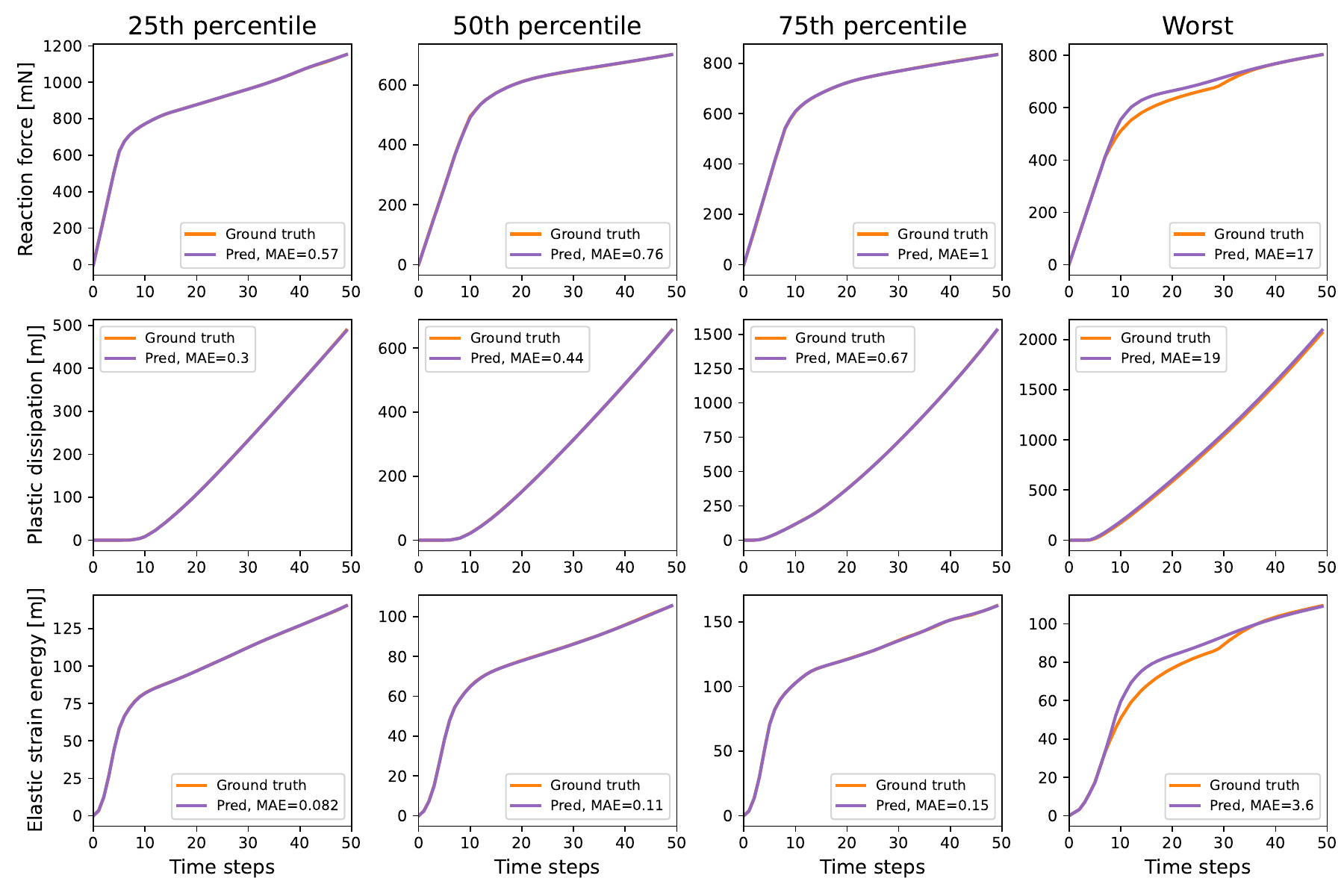} 
    \caption{Comparison of ground truths and GRU predictions for the data set, ranked by percentile of MAE to provide a representative sampling. Here, MAE is ranked independently for each of the four output arrays.}
    \label{prediction_comparison}
\end{center}
\end{figure}

After training the NN, the NN predictions were compared to the ground truths obtained from FE simulations, ranked by the percentile of MAE for each output array. The model with median MAE among the 10 training repetitions was used to generate the plots shown in \fref{prediction_comparison}. The final MAE for this model is 6.07$\times 10^{-3}$.
The amount of data required for training was chosen by checking the loss function value for different percentages of input data. \fref{fig:pct_plot} shows that the loss increases as the percentage of the input data is decreased compared to the reference (80\% data). Hence, we chose 80\% data as input for training. Further, it could be inferred from \fref{fig:loss} that no major overfitting has occurred. The statistical distribution of MAEs is shown in \fref{prediction_comparison}. From the first three columns, up to 75\% percentile, we could see that the GRU model can closely predict the FE simulation results for stress-strain curves, plastic dissipation, and elastic strain energy. Even in the worst case, the GRU model correctly predicts the general shape of the FE-simulated stress-strain curve.

In the current study, the cross-section image of the structure has been parameterized using five design variables. These variables are then used as inputs in the GRU model. Another valid approach is to encode the cross-sectional images of the design via an autoencoder before training the GRU model. This approach was used in the work of He et al. \citep{he2023exploring} for exploring the structure-property relations of thin-walled lattices. However, training the autoencoder can take additional computational resources and is unnecessary when discrete parameter values can parameterize the current design space. Hence in this work, we used the design parameters to describe the designs instead of the autoencoder. However, judging from the comparison with FE data shown in \fref{prediction_comparison}, the prediction accuracy is high even with this simplified approach.

\begin{figure}[h!]
\begin{center}
    \includegraphics[trim={0cm 0cm 0cm 0cm},clip,width=\textwidth]{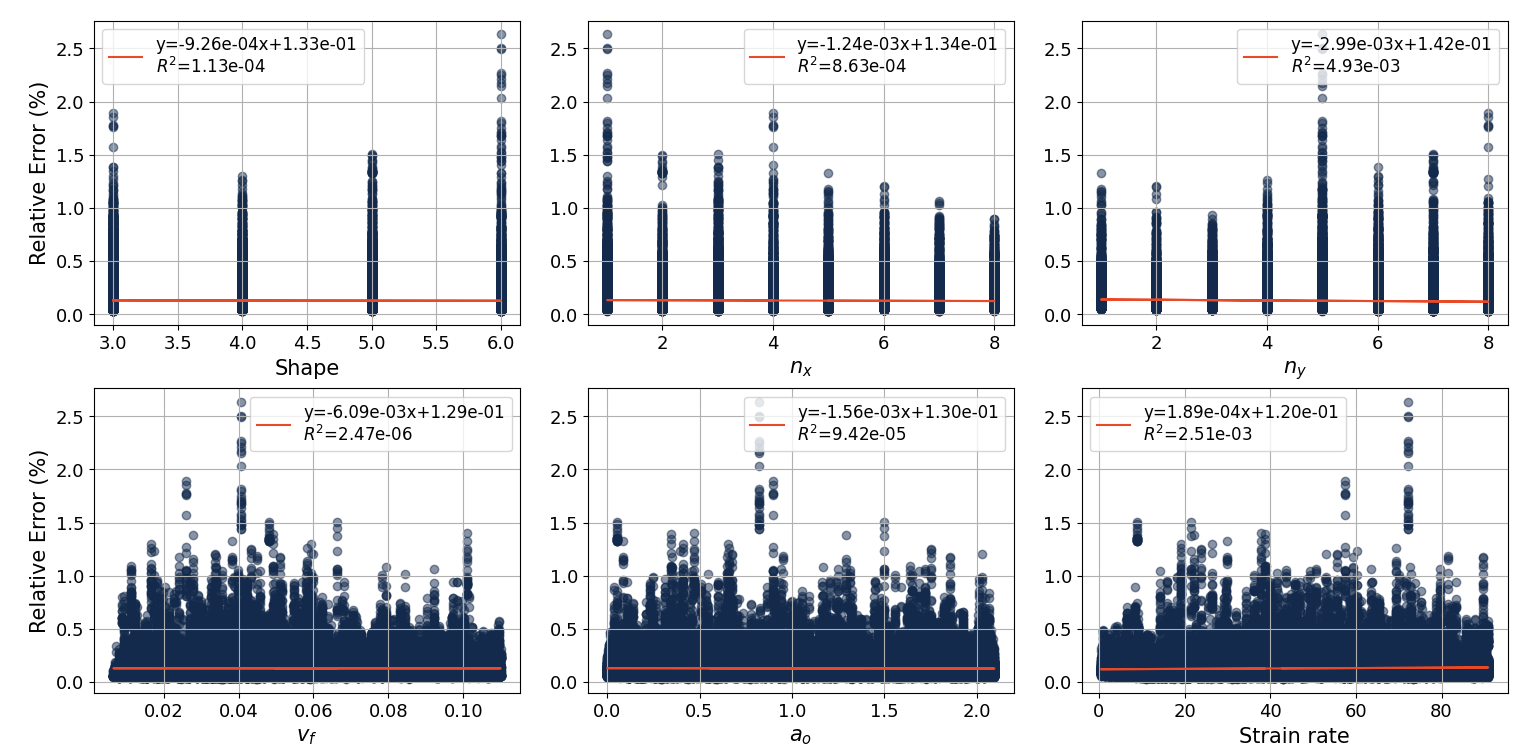} 
    \caption{MAE vs input test data parameters (geometric parameters and loading condition).}
    \label{inputs_worstcase}
\end{center}
\end{figure}

The scatter plots connecting the input data and the corresponding prediction error, along with linear curve fits, were utilized to identify potential correlations. It was observed that, in general, there is no discernible pattern relating the input parameters to the prediction errors, as evidenced by the low $R^{2}$ fitting values. However, a concentration of cases with prediction errors greater than 2\% is observed in \fref{inputs_worstcase}. These cases are concentrated in hexagonal vacancies (shape=6) with one column ($n_{x}$=1) and five rows ($n_{y}$=5), a void volume fraction of 4.1\%, an initial rotation angle of 0.82 radians, and a strain rate of 72.22 $s^{-1}$. However, those cases only constitute 0.0062\% of the total prediction cases.

\subsection{Structure-SEA map}
\label{sec:SEA_map}
\begin{figure}[h!]
\begin{center}
    \includegraphics[scale=0.8,trim = {2.5cm .75cm 2.5cm 2.3cm }, clip, width=0.64\textwidth]{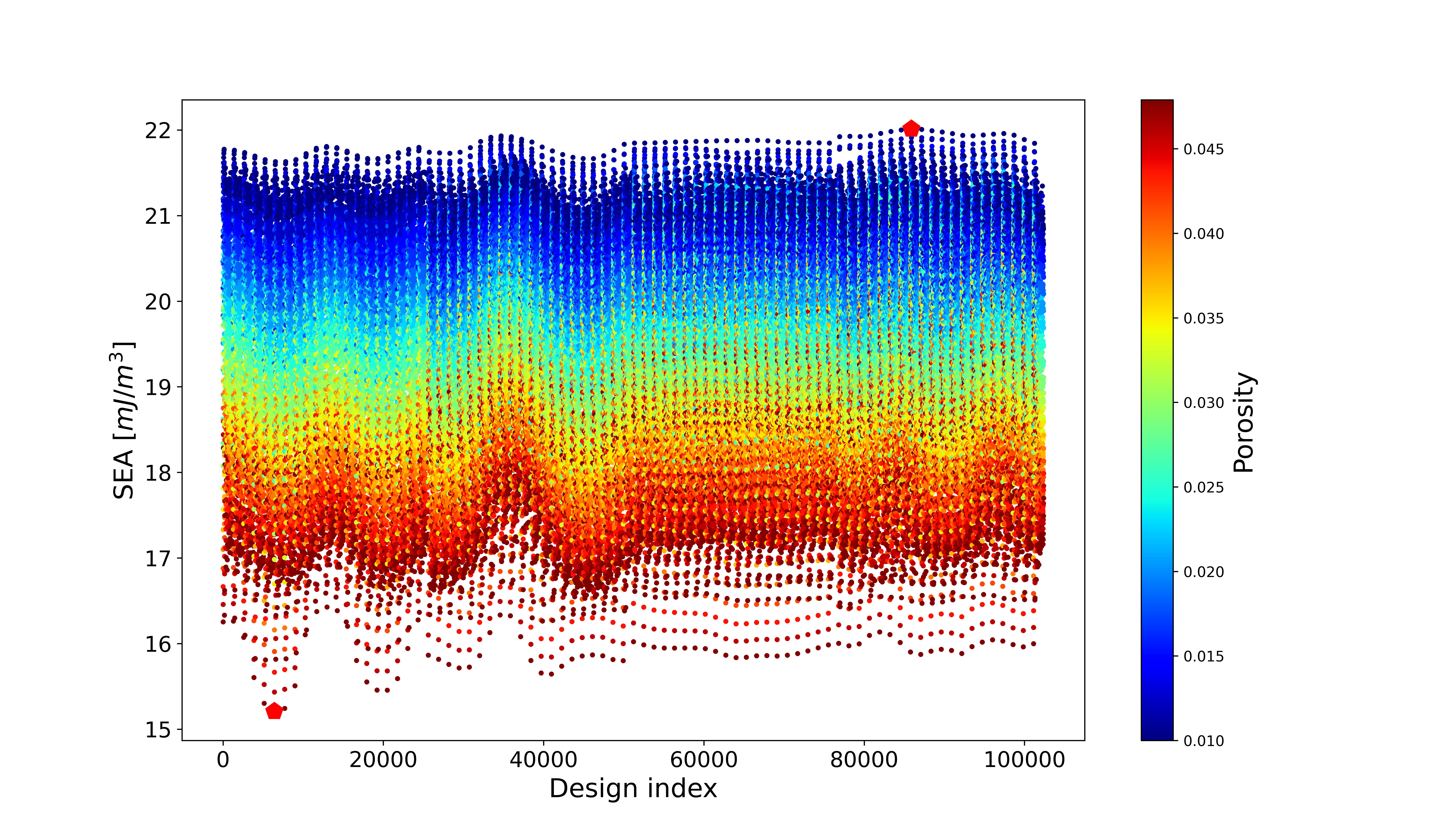} 
    \caption{Structure-property relations at an impact velocity $\rm{v}_{\rm y} = 100 m/s$ and final axial strain of 0.25. The highest and lowest SEA designs are highlighted in solid red pentagons.}
    \label{SEA_map}
    \includegraphics[scale=0.8,trim = {2.5cm .5cm 2.5cm 1.3cm }, clip, width=0.64\textwidth]{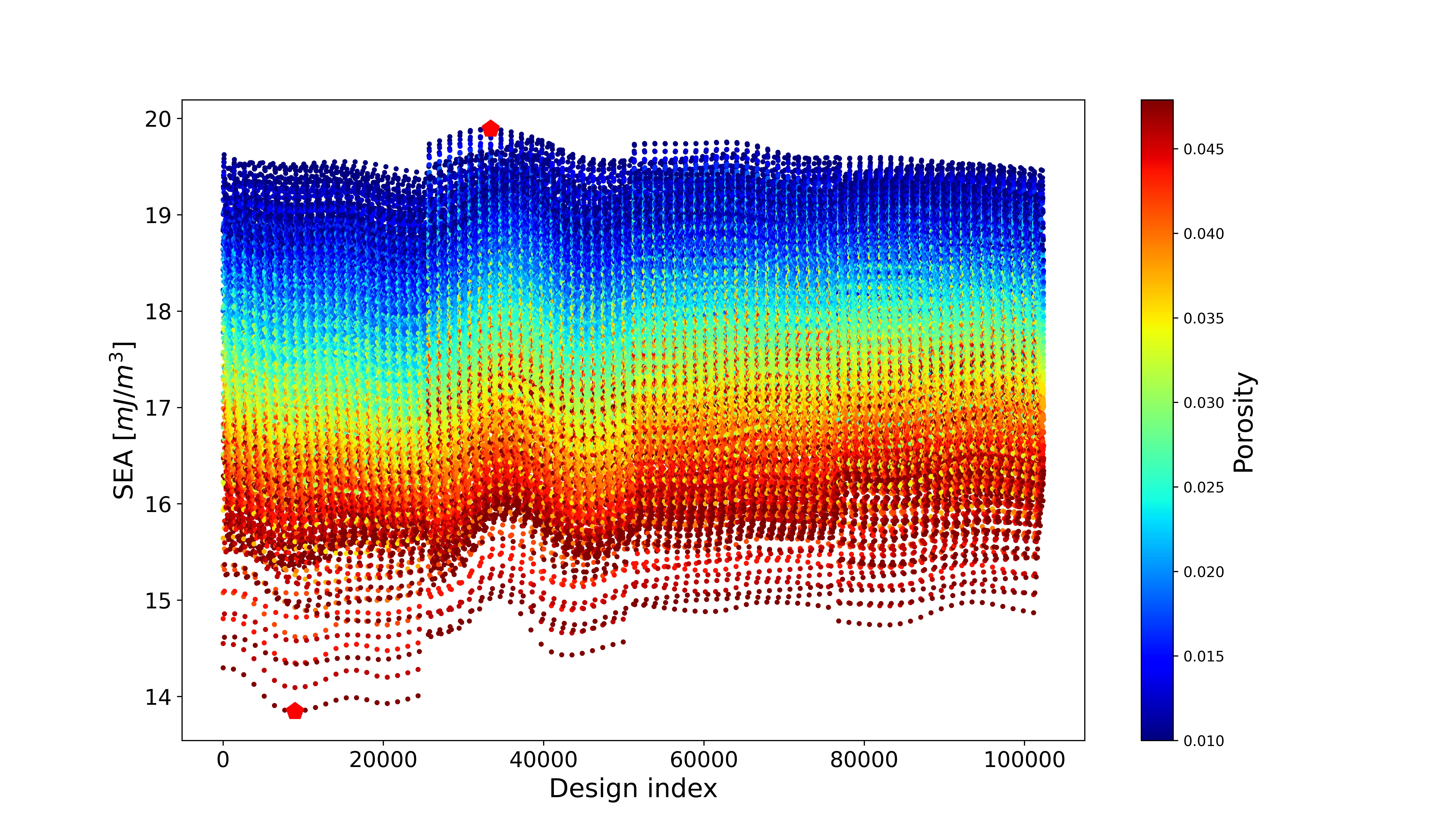} 
    \caption{Structure-property relations at an impact velocity $v_y = 10 m/s$ and final axial strain of 0.25. The highest and lowest SEA designs are highlighted in solid red pentagons.}
    \label{SEA_map_10}
\end{center}
\end{figure}

The Python script described in \sref{sec:opt} was used to calculate SEA from the stress-strain curve predicted by the NN at each design point. Each structure could be represented by a unique design index defined using the first five input parameters to the NN as described in \sref{sec:data_loss}. Finally, the scatter plots for SEA at each design surveyed in the grid search for two different impact velocities are plotted in \fref{SEA_map} and \fref{SEA_map_10}, which show a structure-property map for this chosen design space. 

\begin{figure}[h!]
\begin{center}
    \includegraphics[scale=0.8,trim = {2.5cm .75cm 2.5cm 2.3cm }, clip, width=0.64\textwidth]{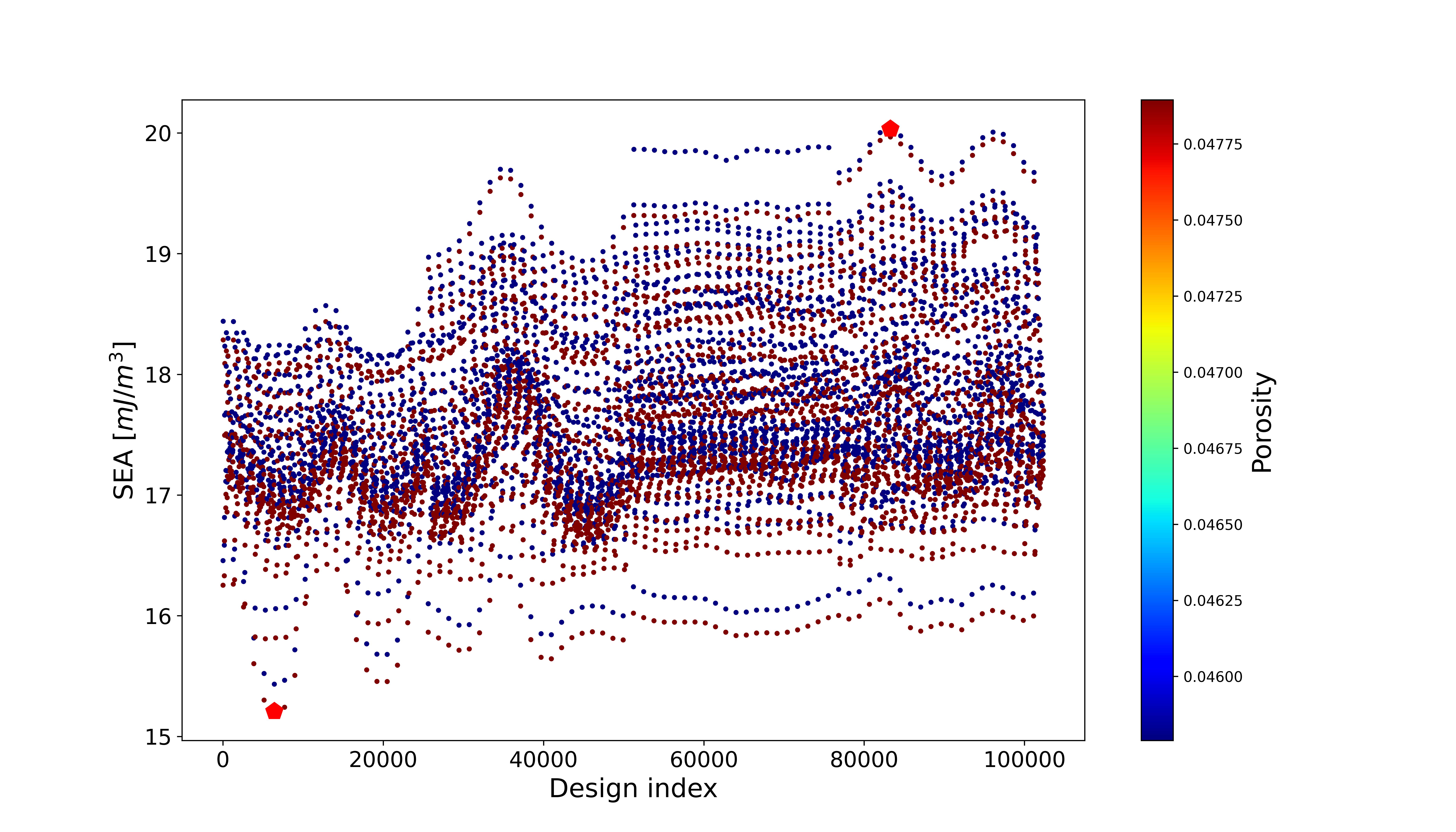} 
    \caption{SEA versus Design Index for porosity volume fraction from 4.5\% to 5\%. The highest and lowest SEA designs are highlighted in solid red pentagons.}
    \label{SEA_4.5_5}
\end{center}
\end{figure}

Using the scatter plot shown in \fref{SEA_map}, we could identify the best and worst designs regarding specific energy absorption within the input design space for the given loading condition and final strain. These two points are also highlighted in the \fref{SEA_map}. Further, the same Python code described in \sref{sec:opt} could be used to plot the SEA for structures with various constraints. For example, \fref{SEA_4.5_5} shows the distribution of SEA for structures with a volume fraction of porosity between 4.5\% and 5\%. 

\subsection{Design trends and observations}
\label{sec:observation}
The structure-energy absorption maps shown in \sref{sec:SEA_map} are useful for obtaining an overview of the entire design space. However, additional design insights could be drawn from the map to guide future design work:
\begin{enumerate}
    \item At the same volume fraction of porosity within the structure, final strain, and indenter velocity exceeding 100 m/s, arranging the pores vertically results in optimal energy absorption. By contrast, the lowest energy absorption is achieved when the porosity is concentrated at the center. The structure illustrated in \fref{Stress_max_SEA} emerged as the most efficient design for SEA, according to the SEA map depicted in \fref{SEA_map}. Conversely, the structure in \fref{Stress_min_SEA} demonstrated the lowest SEA. 
    The structure with maximum SEA (\fref{Stress_max_SEA}) has a porosity of close to 1\%, whereas the one with minimum SEA (\fref{Stress_min_SEA}) has a porosity closer to 5\%. In both instances, we observe a higher stress band that originates at the structure's corners and radiates toward its center during compression. In essence, the presence of material in areas of high stress is crucial for achieving a higher SEA. In the case of the structure in \fref{Stress_max_SEA}, only a few pores are present within the high-stress region. On the other hand, the structure illustrated in \fref{Stress_min_SEA} has its entire porosity at the center, resulting in diminished load-carrying capacity and a lower SEA. 
    \begin{figure}[h!]
    \begin{center}
    \includegraphics[clip,width=.8\textwidth]{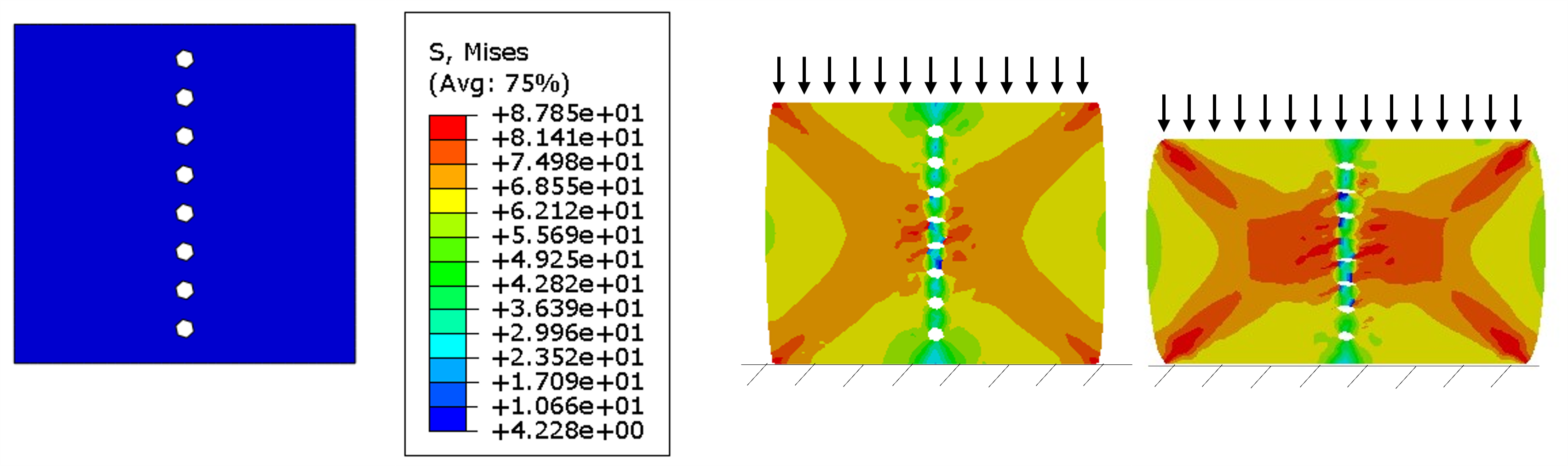} 
    \caption{Initial structure (left) with highest predicted SEA and stress distribution at 12.5\% and 25\% nominal strain (right).}
    \label{Stress_max_SEA}
    \includegraphics[clip,width=.8\textwidth]{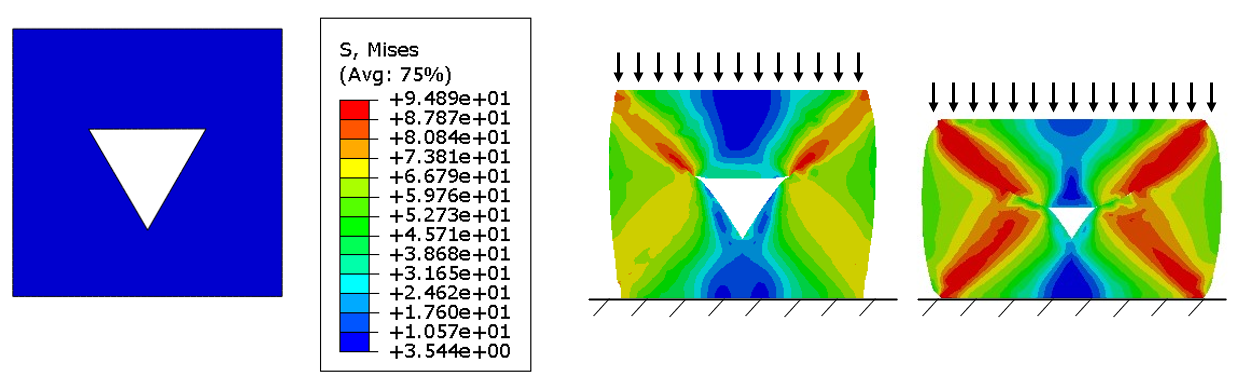} 
    \caption{Initial structure (left) with lowest predicted SEA and stress distribution at 12.5\% and 25\% nominal strain (right).}
    \label{Stress_min_SEA}
    \end{center}
    \end{figure}
    
    \item The orientation of polygonal tubules affects the energy absorption in low-porosity structures. This behavior can be observed in \fref{orientation_trend}, which illustrates two structures with the same square-shaped porosity volume fraction but different angle offsets. When subjected to similar loading conditions, \fref{stack_high_SEA} exhibits 4\% higher energy absorption compared to \fref{stack_low_SEA} as validated by FE simulations. The GRU-predicted trend of how the tubule orientation angle affects the SEA is shown in \fref{trend_orient} for square porosity. The prediction shows a sinusoidal variation, which is reasonable, as the top-down projected load-bearing area (area unaffected by porosity) varies in a sinusoidal fashion.   
    \begin{figure}[h!] 
        \centering
         \subfloat[]{\includegraphics[trim={8cm 0cm 8cm 4cm},clip,width=.2\textwidth]{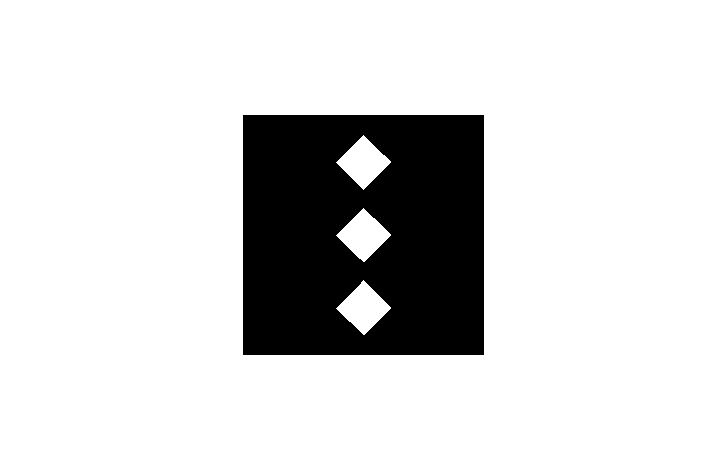}
             \label{stack_low_SEA}
         }
         \subfloat[]{\includegraphics[trim={8cm 0cm 8cm 4cm},clip,width=.2\textwidth]{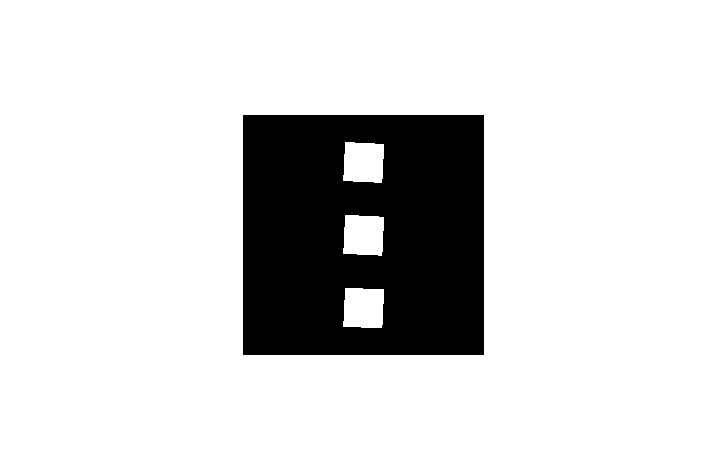}
            \label{stack_high_SEA}
         }
         \subfloat[]{\includegraphics[trim={0cm .cm 0cm 1cm},clip,width=.5\textwidth]{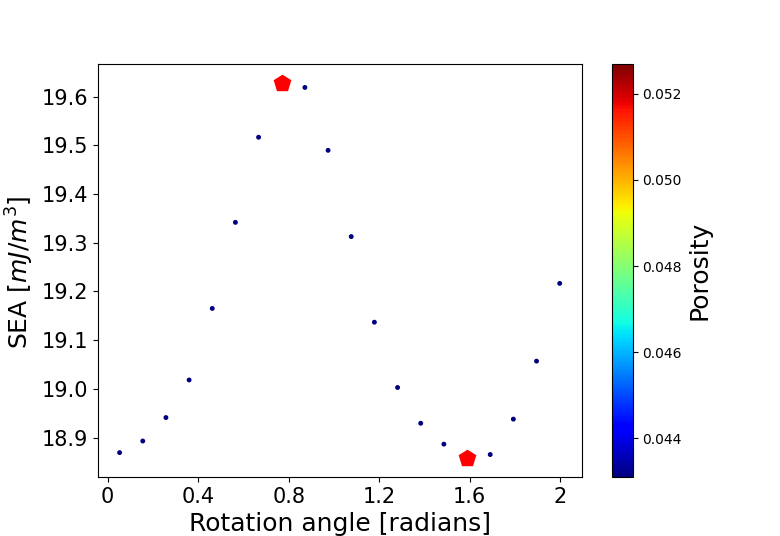}
            \label{trend_orient}
         }
        \caption{Effect of orientation on energy absorption under transverse compression for constant volume fraction: 
        \psubref{stack_low_SEA} Structure absorbing less energy. 
        \psubref{stack_high_SEA} Structure absorbing more energy
        \psubref{trend_orient} Predicted trend as the angle offset is varied.}
        \label{orientation_trend}
    \end{figure}
    
    \item The structure with maximum and minimum SEA depends upon the volume fraction of the porosity. Further, it is also affected by the strain rate and the orientation of polygonal porosity, as shown in \fref{validatation_designs}. For example, the red marks in \fref{SEA_map}, \fref{SEA_map_10}, and \fref{SEA_4.5_5} show different structures (design index) with maximum and minimum SEA.  

    \item The Pearson correlation coefficient is calculated to assess the relationship between SEA and different geometric parameters. Both cases show a strong negative correlation between SEA and volume fraction, indicating that increasing porosity volume fraction generally leads to decreasing SEA. The correlation coefficients for angle offset are close to zero, consistent with the sinusoidal nature of the trend observed in \fref{orientation_trend}. The orientation of pores was found to cause a significant variation in the SEA, with a difference close to 4\%. Hence, in order to obtain correct conclusions using correlation analysis, it is necessary to employ exploratory grid search methods to identify select designs that exhibit a high SEA. The number of pores in the x-direction is negatively correlated to SEA, while the number of pores in the y-direction is positively correlated. Apart from that, a minor correlation is observed for other variables. 
    \begin{figure}[h!]
    \begin{center}
        \includegraphics[scale=0.9, trim={0cm .cm 0cm 1.3cm},clip, width=0.64\textwidth]{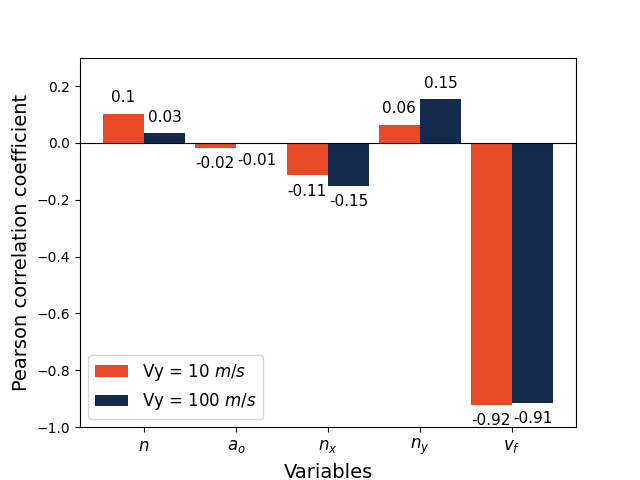} 
        \caption{Pearson correlation coefficient between SEA and geometric variables at two different indenter velocities.}
        \label{pearson_co}
    \end{center}
    \end{figure}
\end{enumerate}

\section{Conclusions and future work}
\label{sec:conc}

In this work, a parametric framework was developed to generate bio-inspired low-porosity designs with tubules of various shapes, orientations, and in-plane arrangements. The structures were made from PC-ABS with rate-dependent elastoplastic behavior. FE simulations were conducted to obtain the stress-strain curves of the structures at different impact velocities during transverse loading. Using the FE simulation data, a GRU model was trained to predict the stress-strain curve for low-porosity bio-inspired structures under dynamic transverse compression loading. Data augmentation techniques were implemented to reduce the number of simulations required using Abaqus.

The trained NN model could make accurate predictions (MAE: 6.07 $\times 10^{-3}$) for SEA of all the structures across a range of final strains and strain rates. Further, the trained neural network was used to survey the entire design space with 128000 structures at each strain rate. Overall, the trained model NN was able to generate all the performance predictions extremely efficiently, even on low-end laptops. The stress-strain response for each structure could be predicted in 0.16ms. Hence, it renders itself a suitable guide in preliminary design stages to quickly survey designs for more detailed analyses. 

Using the predictions of the trained NN, key observations were made and summarized below:

\begin{itemize}
   \item The SEA maps generated using grid search based on geometric variables facilitated the identification of several design trends obtained from the trained NN model.
   \item Our study delved deep into the influences of porosity arrangement, volume fraction, strain rate, and orientation on the SEA. Two standout findings were:
   \begin{itemize}
     \item Varying the orientations of pores can result in approximately 4\% difference in SEA.
     \item Vertical arrangement of pores at the same volume fraction led to greater SEA. 
   \end{itemize}
   \item The SEA maps, beyond their immediate application, contribute to understanding the effect of geometric parameters on SEA under varying loading conditions.
   \item The Pearson correlation analysis augmented the study by drawing connections between different geometric parameters and the SEA.
   \begin{itemize}
     \item The results indicated a strong negative correlation between SEA and porosity volume fraction, while minor correlations were observed for other variables.
     \item The minor correlation between the variables reinforces the need to utilize exploratory grid searches to identify select configurations that exhibit higher SEA under given loading conditions.
   \end{itemize}
 \end{itemize}

In future work, gradients of the GRU model could be utilized to define an inverse design problem and generate new designs. In the current work, periodic boundary conditions were not enforced on the representative volume when comparing different structures. The effect of enforcing periodic boundary conditions could also be explored in future work. Lastly, the structures analyzed in this study using FEA were not compared against the experimental results, a key limitation of the current work. Hence, an experimental validation of the FE simulation model would provide further insights into the model accuracy and the energy absorption capabilities of the low-porosity structures.

\section*{Data availability}
The data and source code that support the findings of this study will be available upon  request during the  review process, and it will be open source after the publication is online.

\section*{Conflict of interest}
The authors declare that they have no conflict of interest.

\section*{Acknowledgements}
We acknowledge the support of the National Science Foundation grant (MOMS-1926353) and the Army Research Office contract (No. W 911NF-18-2-0067).

\section*{CRediT author contributions}
\textbf{Shashank Kushwaha}: Conceptualization, Methodology, Software, Formal analysis, Investigation, Data Curation, Writing - Original Draft. \textbf{Junyan He}: Methodology, Software, Formal analysis, Writing - Original Draft. \textbf{Diab Abueidda}: Supervision, Writing - Review \& Editing. \textbf{Iwona Jasiuk}: Supervision, Resources, Writing - Review \& Editing, Funding Acquisition.

\bibliographystyle{unsrtnat}
\setlength{\bibsep}{0.0pt}
{\scriptsize \bibliography{References.bib} }
\end{document}